\documentclass[%
 reprint,
superscriptaddress,
 amsmath,amssymb,
 aps,
 prl
]{revtex4-1}
\usepackage{upgreek}
\usepackage{graphicx}% Include figure files
\usepackage{dcolumn}% Align table columns on decimal point
\usepackage{bm}% bold math
\begin{document}

\preprint{APS/123-QED}

\title{Coherent continuous wave terahertz spectroscopy using Hilbert transformation}%Response analysis of linear systems in coherent frequency domain terahertz spectroscopy using Hilbert transformation}% Force line breaks with \\
%\thanks{A footnote to the article title}%

\author{Dominik Walter Vogt}

 \email{d.vogt@auckland.ac.nz}
 %\altaffiliation[Also at ]{Physics Department, XYZ University.}%Lines break automatically or can be forced with \\
\author{Miro Erkintalo}%
\author{Rainer Leonhardt}%

\affiliation{Department of Physics, The University of Auckland, Auckland 1010, New Zealand}%
\affiliation{The Dodd-Walls Centre for Photonic and Quantum Technologies, New Zealand}%
%\collaboration{MUSO Collaboration}%\noaffiliation

%\author{Charlie Author}
% \homepage{http://www.Second.institution.edu/~Charlie.Author}
%\affiliation{
% Second institution and/or address\\
% This line break forced% with \\
%}%
%\affiliation{
% Third institution, the second for Charlie Author
%}%
%\author{Delta Author}
%\affiliation{%
% Authors' institution and/or address\\
% This line break forced with \textbackslash\textbackslash
%}%

%\collaboration{CLEO Collaboration}%\noaffiliation

\date{\today}% It is always \today, today,
             %  but any date may be explicitly specified

\begin{abstract}
Coherent continuous wave (CW) terahertz spectroscopy is an extremely valuable technique that allows for the interrogation of systems that exhibit narrow resonances in the terahertz (THz) frequency range, such as high-quality (high-Q) THz whispering-gallery mode resonators. Unfortunately, common implementations are dramatically impaired by deficiencies in the used data analysis schemes. Here, we show that the physics of the problem presents an elegant solution whose full potential has remained overlooked until now. We argue that, thanks to the causality of physical systems, Hilbert transformation can be used to analyze the frequency response of linear systems with arbitrarily narrow resonance features in coherent CW THz spectroscopy. In particular, by establishing that signals encountered in typical experiments are closely related to analytic signals, we demonstrate that Hilbert transformation is applicable even when the envelope varies rapidly compared to the oscillation period. %Our work demonstrates that Hilbert transformation is an extremely powerful tool for the study of high quality resonances in the THz frequency range using continuous wave THz spectroscopy.
\end{abstract}

\pacs{Valid PACS appear here}% PACS, the Physics and Astronomy
                             % Classification Scheme.
%\keywords{Suggested keywords}%Use showkeys class option if keyword
                              %display desired
\maketitle

Hilbert transformation provides a compelling tool to retrieve the instantaneous amplitude (envelope) and phase of an oscillating signal \cite{liu2012hilbert}. However, common wisdom asserts that the transformation is only applicable to signals with a slowly varying envelope compared to the oscillation period \cite{kong2018high}. The prerequisite of a slowly varying envelope thus seemingly precludes the desirable application of the Hilbert transformation to generic narrow resonant features modulated on an oscillating signal. Such applications arise, for example, in coherent CW THz spectroscopy of gases or artificial high-Q structures \cite{hepp2016cw,vogt2018ultra,mouret2006anomalous}.

Coherent CW THz spectroscopy is an extremely powerful technique that can yield a wealth of information hidden from traditional methods, and has revolutionized fields like sensing and material characterization \cite{tonouchi2007cutting,bigourd2006detection,ferguson2002materials,liu2007terahertz,mittleman2013sensing,roggenbuck2010coherent,dexheimer2007terahertz,siegel2004terahertz,jansen2010terahertz,lee2009principles}. In particular, CW THz spectroscopy -- as opposed to time-domain THz spectroscopy -- provides the advantage of compact, cost effective systems with superior frequency resolution \cite{karpowicz2005comparison}. The signal measured by these systems is a photocurrent that exhibits sinusoidal oscillation with frequency: the desired spectral information is extracted by numerically retrieving the envelope and phase of the signal. Traditionally, this has been achieved by cumbersomely inspecting the extrema of the oscillating signal, leading to an effective frequency resolution that is usually one or 
two orders of magnitude lower than the actual frequency step size of the acquired signal \cite{Vogt:17}. Recent works have shown that this issue -- which is severely impacting the capabilities of CW THz spectroscopy -- can be overcome by taking advantage of the Hilbert transformation \cite{Vogt:17,xiao2018high,kong2018high}. However, in compliance with common wisdom, it has been pointed out that Hilbert transformation is limited to CW THz spectroscopy of systems with broad resonance features compared to the oscillation period of the measured signal \cite{kong2018high}.         

In this Letter, we show that this presumed limitation is in fact incorrect, and we establish for the first time a universal, comprehensive foundation for the applicability of Hilbert transformation for CW THz spectroscopy. Our analysis reveals that the coherent signal interrogated in the frequency domain experiments is analytic, i.e. its real and complex parts are related via Hilbert transformation. Consequently, applying the Hilbert transformation to the measured signal, which is directly proportional to the real part of the analytic signal, provides the complete information about the frequency response of the investigated system. We experimentally and numerically demonstrate this concept for three THz systems where narrow resonances manifest themselves: whispering-gallery mode resonators (WGMRs), a photonic crystal cavity waveguide (PCCW) \cite{yee2009high,bingham2008terahertz,chen2014terahertz}, and a planar split-ring resonator (SRR) metamaterial \cite{chen2009metamaterial,chen2006active,
choi2011terahertz,tao2008highly}. For each of these systems -- which correspond to active research areas in their own right -- we show that the Hilbert transformation allows for the analysis of resonance features regardless of whether they vary slowly with respect to the oscillating signal or not. We expect our findings to significantly enhance the capabilities of CW THz spectroscopy. 

%Furthermore, we show that the interrogated signal contains unimpaired information about the investigated system.

%For any linear system, the time response ${E}^{\textrm{THz}}_{\textrm{out}}(t)$ is given as the convolution of the transfer function $S(t)$ and the input (carrier) signal ${E}^{\textrm{THz}}_{\textrm{in}}(t)$: 
%\begin{equation}
%{E}^{\textrm{THz}}_{\textrm{out}}(t)=S(t)\ast{E}^{\textrm{THz}}_{\textrm{in}}(t).
%\label{eq:1}
%\end{equation}

%Our scheme not only allows the characterization of arbitrarily narrow resonance features compared to the oscillation frequency, but also introduces a significant advantage in both frequency resolution and/or acquisition time for CW THz spectroscopy in general \cite{Vogt:17}.

We consider a situation where a THz field ${E}_{\textrm{in}}(t)$ interacts with a linear system described by a time response function $S(t)$. The output field is given in the frequency domain by the expression:\begin{equation}
{\tilde{E}}_{\textrm{out}}(\omega)=\tilde{S}(\omega){\tilde{E}}_{\textrm{in}}(\omega),
\label{eq:1}
\end{equation}
where ${\tilde{E}}_{\textrm{out}}(\omega) = \mathcal{F}[{E}_{\textrm{out}}(t)]$ is the Fourier transform of ${E}_{\textrm{out}}(t)$ (similar notation applies to the other two variables).

For physical systems, the time response ${E}_{\textrm{out}}(t)$ is causal. Accordingly, the frequency response ${\tilde{E}}_{\textrm{out}}(\omega)$ is an analytic signal \cite{stearns1976digital}. This implies that the real and imaginary parts of $\tilde{E}_{\textrm{out}}(\omega)$ are linked via the Hilbert transformation $\mathcal{H}$ \footnote{Note that in the physics context Hilbert transformation is known as the Kramers-Kronig relation, and is closely related to concepts of single-sideband modulation.}:
\begin{equation}
{\tilde{E}}_{\textrm{out}}(\omega)=\operatorname{Re}[{\tilde{E}}_{\textrm{out}}(\omega)]+i\mathcal{H}\{\operatorname{Re}[{\tilde{E}}_{\textrm{out}}(\omega)]\}.
\label{eq:2}
\end{equation}
%From Eqs. (\ref{eq:1}) and (\ref{eq:2}), it is evident that applying the Hilbert transformation to the photocurrent allows to retrieve the unimpaired complex frequency response of the system under study.

In typical coherent CW THz spectroscopy experiments, one detects a photocurrent that is directly proportional to the real part of the frequency response $\tilde{E}_{\textrm{out}}(\omega)$. From Eq. (\ref{eq:2}), it is evident that applying the Hilbert transformation to the photocurrent allows to retrieve the unimpaired complex frequency response of the system under study. Assuming the input THz field ${\tilde{E}}_{\textrm{in}}(\omega)$ is known, Eq. (\ref{eq:1}) then yields the complete frequency-domain response function $\tilde{S}(\omega)$, which contains all the spectroscopic information of the system. As all physical systems are causal, these arguments hold regardless of the specific form of the transfer function $\tilde{S}(\omega)$ or ${\tilde{E}}_{\textrm{in}}(\omega)$. Consequently, Hilbert transformation can be used to characterize the frequency response of any physical linear system with coherent CW spectroscopy, especially for the desirable case of resonance features arbitrarily narrow compared to 
the oscillation period of the interrogated signal.

To exemplify the strength of the Hilbert transformation approach, we next consider three illustrative examples. First, we focus on THz WGMRs, which show unprecedented Q-factors and provide exciting opportunities for the THz frequency range \cite{vogt2018ultra,Vogt:172,vogt2018thermal,Preu:08,xie2018terahertz}. Here the THz field ${E}_{\textrm{in}}(t)$ is coupled into a WGMR using the evanescent field of a single-mode waveguide. The frequency response of this system is known, and it can be written as \cite{Gorodetsky:99}:          
\begin{equation}
{\tilde{E}}_{\textrm{out}}(\omega)=\bigg[1-\frac{2 {\delta}_{\textrm{c}}}{\delta_0+{\delta}_{\textrm{c}}-i ({\omega}-{\omega}_{0})}\bigg]{\tilde{E}}_{\textrm{in}}(\omega),
\label{eq:3}
\end{equation}
where ${\delta}_{0}$ and ${\delta}_{\textrm{c}}$ are the intrinsic loss rate and the coupling rate, respectively, and ${\omega}_{0}=2\pi {f}_{0}$ is the (angular) resonance frequency. Like for all CW THz spectroscopy experiments, ${\tilde{E}}_{\textrm{in}}(\omega)$ reads: 
\begin{equation}
{\tilde{E}}_{\textrm{in}}(\omega)={E}_{0} {e}^{i \frac{\Delta L}{c} {\omega}}.
\label{eq:4}
\end{equation}
Here, $c$ is the speed of light, ${E}_{0}$ is the THz field amplitude, and $\Delta L$ is the optical path difference between the emitter and detector arms of the spectrometer including the THz path \cite{1367-2630-12-4-043017}. %Varying $\Delta L$ changes the carrier frequency ${\textrm{f}}_{c}=c/\Delta L$ of the THz signal. Since the THz carrier signal relevant to this work is already in the frequency domain we define the carrier frequency as the frequency spacing between two subsequent maxima of the sinusoidally-varying, frequency-dependent THz signal (photocurrent). Please note that while it is possible to some extent to increase $\Delta L$ (and therefore decrease the carrier frequency) in the experiment, this is highly undesirable as it increases the noise and complexity of the experiment \cite{roggenbuck2012using}.

\begin{figure}[t]
\centering
\includegraphics[width=\linewidth]{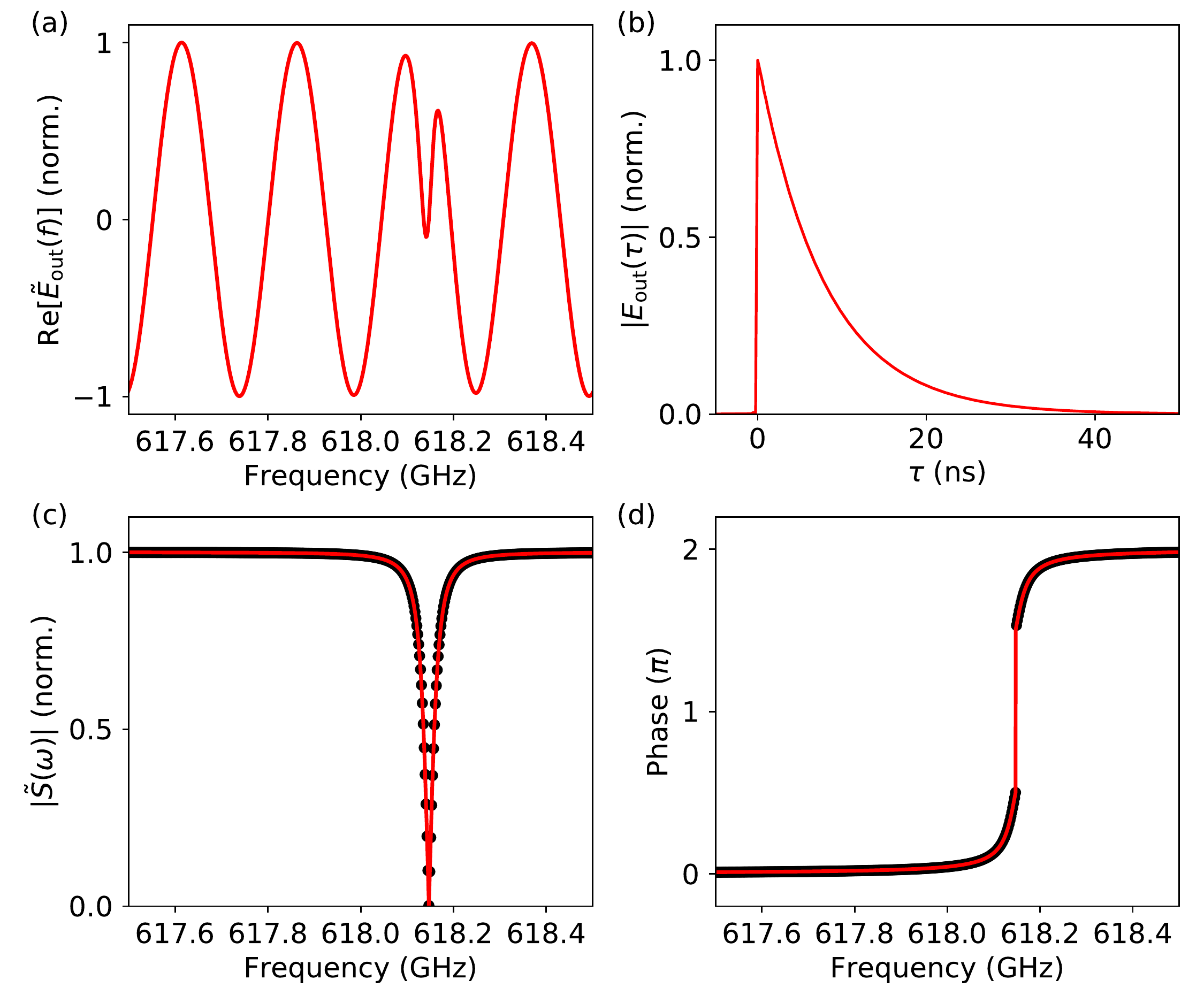}
\caption{(a) Real part of Eq. (\ref{eq:3}). (b) Envelope of the time response. (c) Comparison of $|\tilde{S}(\omega)|$ retrieved from the analytical model (cf. Eq. (\ref{eq:3}), red solid line) and Hilbert transformation (black circles). (d) Comparison similar as in (c) but for the phase of the frequency response function.}
\label{fig:1}
\end{figure}

We consider a critically-coupled WGMR similar to the experiments that will follow, with ${f}_{0}=618.147\,\textrm{GHz}$ and a Q-factor of $1.5\times10^4$ ($\delta_0 = {\delta}_{\textrm{c}}=65\,\mathrm{MHz}$). The analytical photocurrent [real part of Eq. (\ref{eq:3})] shown in Fig. \ref{fig:1}(a) is an oscillating function with $\omega$, onto which the resonance feature is imprinted. Because common wisdom has it that Hilbert transformation is only applicable to oscillating signals with a slowly varying envelope compared to the underlying period, one might expect that Hilbert transformation is only valid if the photocurrent oscillates much faster than the width of the resonance feature \cite{kong2018high}. This is incorrect: because the measured photocurrent is the real part of an analytical signal, applying the Hilbert transformation allows to retrieve the full spectroscopic information.

The fact that the photocurrent is the real part of an analytical signal can be readily verified by establishing the causality of the corresponding time response ${E}_{\textrm{out}}(t) = \mathcal{F}^{-1}\{{\tilde{E}}_{\textrm{out}}(\omega)\}$, which is explicitly given by:
\begin{equation}
%{\mathcal{F}}^{-1}\Big\{{\tilde{E}}_{\textrm{out}}(\omega)\Big\}
{E}_{\textrm{out}}(\tau)=\sqrt{2 \pi}\Big[\delta(\tau)-2{\delta}_{\textrm{c}}{e}^{-\tau({\delta}_{\textrm{c}}+{\delta}_{\textrm{o}})}\Theta(\tau)\Big]{e}^{-i{\omega}_{0}t}.
\label{eq:5}
\end{equation}
Here, $\delta(\tau)$ is the Dirac delta function, $\Theta(\tau)$ is the Heaviside step function, and $\tau = t-{t}_{c}$ with ${t}_{c}=\Delta L/c$. A typical time response is visualized in Fig. \ref{fig:1}(b), where we plot $|{E}_{\textrm{out}}(\tau)|$ for our resonator parameters (Dirac impulse removed for clarity). Equation (\ref{eq:5}) is also commonly known as the ring-down signal of the WGM \cite{armani2003ultra}.     

%Next, we compare the corresponding envelopes and phases obtained from both the analytical model based on Eq. (\ref{eq:3}) (red solid line) and the Hilbert transformation applied to the photocurrent (black circles) in Fig. \ref{fig:1}(c) and (d), respectively. 

In Figs. \ref{fig:1}(c) and (d), we respectively compare the envelopes and phases of $\tilde{S}(\omega)$ obtained from the analytical model based on Eq. (\ref{eq:3}) and the Hilbert transformation applied to the photocurrent. As expected, the calculated envelopes and phases are identical, highlighting how the Hilbert transformation approach indeed allows for the full reconstruction of the system's frequency response. We must emphasize that the amplitude full-width half-maximum (FWHM) of the WGM (24\,MHz) is about ten times smaller compared to the 250\,MHz oscillation period of the photocurrent, underlining the fact that Hilbert transformation can readily be used to analyze THz WGMs much narrower compared to the oscillation period. %Note that here we define the oscillation frequency as the frequency spacing between two subsequent maxima of the sinusoidally varying frequency dependent photocurrent. 

\begin{figure}[t]
\centering
\includegraphics[width=\linewidth]{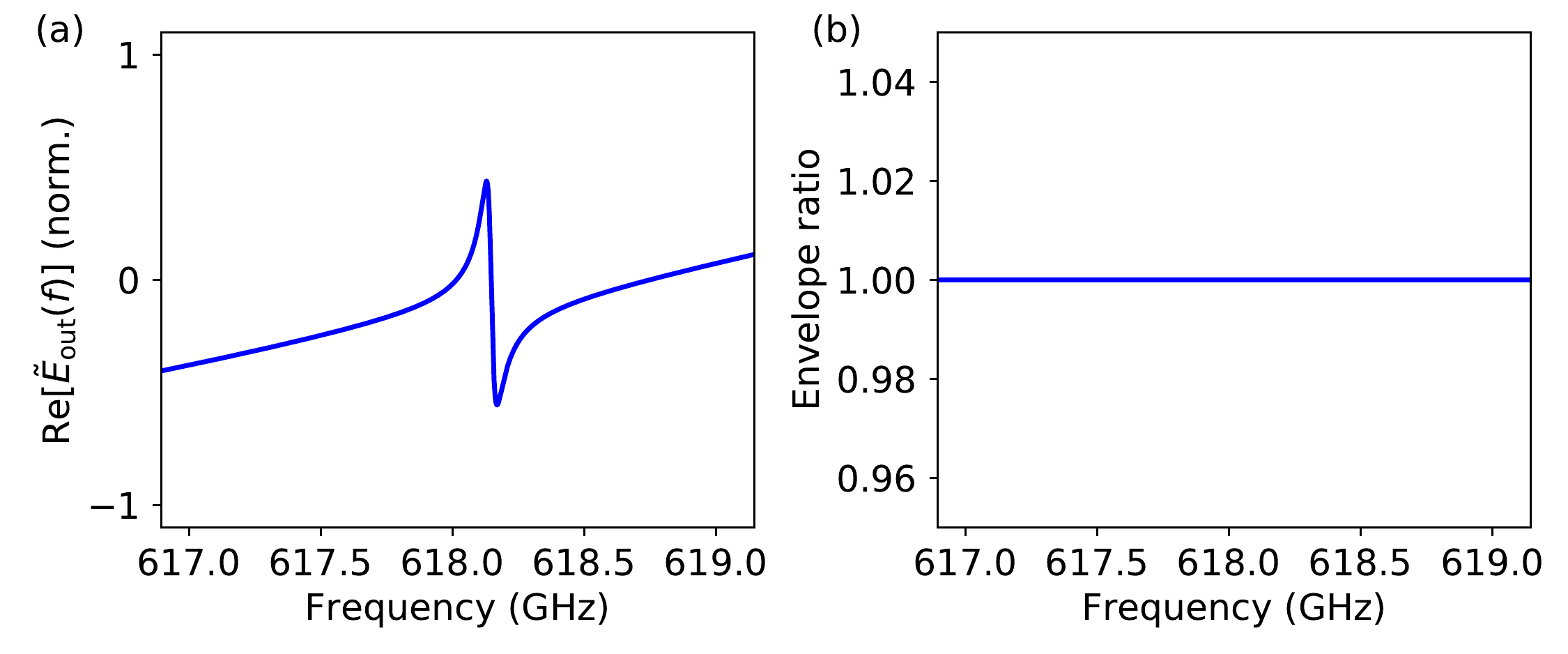}
\caption{(a) Real part of Eq. (\ref{eq:3}) for the THz WGM shown in Fig. \ref{fig:1}, but with a 100 times larger oscillation period of the photocurrent. (b) Ratio of the envelopes calculated with Hilbert transformation and the analytical model.} 
\label{fig:2}
\end{figure}

For further demonstration, Fig. \ref{fig:2}(a) shows the real part of Eq. (\ref{eq:3}) (blue solid line) for the same THz WGM at critical coupling but with a 100 times larger oscillation period (25\,GHz). The ratio of the corresponding envelopes calculated with the Hilbert transformation and from the analytical model is unity [see Fig. \ref{fig:2}(b)]. The amplitude FWHM of the resonance is more than a thousand times smaller compared to the photocurrent oscillation period, yet the Hilbert transformation allows for the full recovery of the frequency response. Of course, numerically the Hilbert transformation works best when a sufficiently large integer number of oscillations of the photocurrent are analyzed.

%Furthermore, for satisfactory performance, a sufficiently large frequency range around the resonance feature must be sampled to avoid edge artifacts. 

%However, numerically, care has to be taken with edge effects from the Hilbert transformation, i.e. a sufficiently large frequency range around the resonance has to be sampled. This is demonstrated with the black solid line shown in Fig. \ref{fig:2}(a) which is identical to the orange solid line but sampled over a narrower frequency range (two instead of nine cycles). Where the orange solid line is identical to the curve in Fig. \ref{fig:1}. The corresponding ratio shows deviations from unity at the resonance frequency [see black solid line in Fig. \ref{fig:2}(b)]. For comparison the ratio calculated for the orange solid line from Fig. \ref{fig:2}(a) with a wider sampling range is also shown (orange dashed line). %As a general guideline at least three to four FWHMs should be sampled in order to minimize the unwanted edge effects from the Hilbert transformation.

%As expected, the envelope retrieved from the analytical model is identical to the previous curve (orange lines). However, the envelope obtained from the Hilbert transformation shows considerable artifacts (black dashed lines). As a general guideline at least three to four carrier signal oscillations should be sampled in order to minimize the unwanted edge effects from the Hilbert transformation.

We now show that the Hilbert transformation approach can be applied to experimental data. The system under study is a spherical WGMR with a diameter of 4\,mm made of high-resistivity float-zone grown silicon (HRFZ-Si) \cite{vogt2018thermal}. The THz WGMs are excited using the evanescent field of a single-mode air-silica step index waveguide. The measurements are performed with a standard CW THz spectroscopy setup based on heterodyne detection (Toptica TeraScan 1550 nm) \cite{Deninger2015}. The experimental setup is described in detail in a previous publication \cite{vogt2018ultra}.

\begin{figure}[t]
\centering
\includegraphics[width=\linewidth]{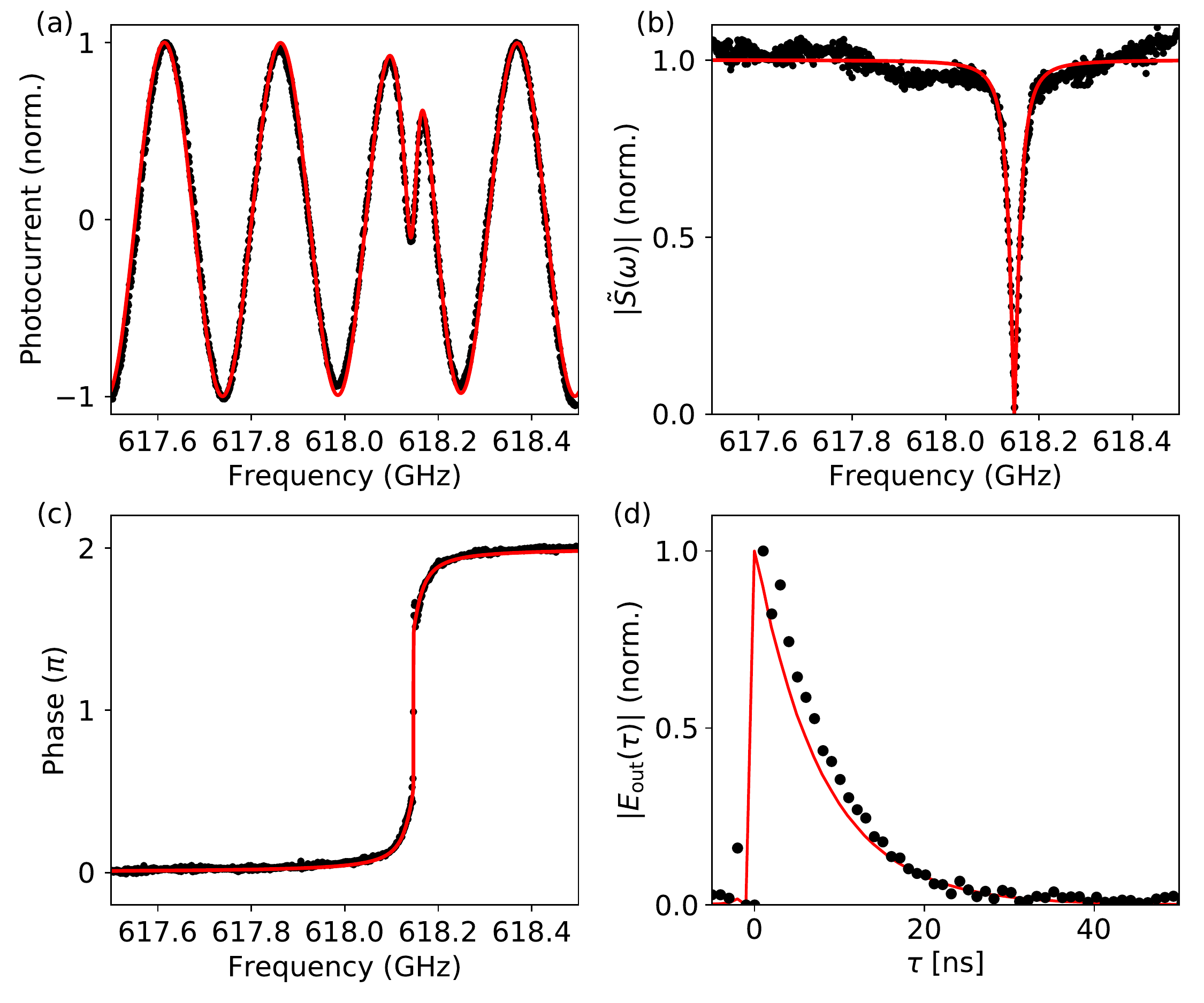}
\caption{Measured (black dots) and fitted (red solid line) photocurrent according to Eq. (\ref{eq:3}). (b) and (c) compare the envelopes and phase profiles of the frequency response function calculated from the Hilbert transformation applied to the photocurrent and the fit, respectively. (d) Shows the corresponding envelopes of the time response.} %The fitting parameters are: ${E}_{0}^{\textrm{THz}}=2.75\,\textrm{nA}, \Delta L=1.20\,\textrm{m}, \Phi=-2\pi, \delta_0=63.0\,\textrm{MHz}, {\delta}_{\textrm{c}}=63.4\,\textrm{MHz}, {f}_{0}=618.15\,\textrm{GHz}$.}
\label{fig:3}
\end{figure}

The measured photocurrent is shown in Fig. \ref{fig:3}(a) in the frequency range from 617.5\,GHz to 618.5\,GHz with black dots. The red solid line shows the fit of the real part of Eq. (\ref{eq:3}) to the experimental data. Similar to the numerical example above, we apply the Hilbert transformation on the experimentally measured photocurrent to reconstruct the analytic signal ${\tilde{E}}_{\textrm{out}}(\omega)$. From this, we can fully reconstruct the frequency response function $\tilde{S}(\omega)$ of the system. In Figs. \ref{fig:3}(b) and \ref{fig:3}(c), we compare the envelope and phase of the frequency response function with corresponding profiles obtained from a fit of the experimental data on the real part of Eq. (\ref{eq:3}). The agreement is excellent, revealing the amplitude linewidth of the WGM to be 25\,MHz. Figure \ref{fig:3}(d) shows corresponding comparison of the time response (obtained by inverse Fourier transforming the experimentally reconstructed analytic signal), and again we see 
outstanding agreement. In particular, it is remarkable that the detection scheme in the CW THZ spectroscopy experiment and experimental imperfections do not impact the interrogated signal of the system according to Eq. (\ref{eq:3}). We must note that, in general, standing waves present in the setup, a frequency-dependent performance of the employed THz emitter/detector, and a frequency-dependent optical path length $L$ (oscillation frequency), render fitting of the photocurrent with an analytical model impractical as well as inaccurate. The Hilbert transformation is unaffected by such issues, and therefore offers a superior method in general compared to analytical fitting. Furthermore, using Hilbert transformation, any deviations are readily eliminated by evaluating the frequency response of an investigated sample compared to the frequency response of an appropriate reference scan, as is common for THz spectroscopy.

The WGMR example considered above is a special case for which the functional form of the frequency response is known [cf. Eq. (\ref{eq:3})]. To highlight the versatility of the Hilbert transformation approach, we now consider two scenarios where the functional form of the frequency response function $\tilde{S}(\omega)$ is not known. Specifically, we use COMSOL Multiphysics\textsuperscript{\textregistered} to numerically extract the complex frequency response of a photonic crystal waveguide and a split-ring resonator metamaterial, and subsequently generate mock photocurrent data by multiplying the appropriate simulated frequency response function with Eq. (\ref{eq:4}). For demonstration purposes, the oscillation period of the photocurrent has been chosen to be significantly higher than the FWHM of the resonances.

%The presented simulations not only allow conclusions about the applicability of the Hilbert transformation to two very common artificial structures, but also demonstrate how to verify the validity of the approach for arbitrarily complicated structures where an analytical solution is not necessarily known. 

%Please note that the structures are not optimized for very high-Q, and are merely representatives for those two common structures. Furthermore,

%In the first step, the simulations are used to extract the complex frequency response $S(\omega_\mathrm{THz})$ (S-parameters) of the structures under test. Subsequently, mock data is generated by taking the appropriate complex S-parameter multiplied with a carrier signal (see Eq. \ref{eq:4}). Finally, we examine whether the time response, envelope and phase extracted from the mock data agree with the results obtained with the Hilbert transformation applied to the mock data.

%To highlight the versatility of the Hilbert transformation approach, we next discuss the numerical analysis of a photonic crystal cavity waveguide (PCCW) and a planar split-ring resonator metamaterial (SRRs). The frequency-domain finite-element method simulations are performed with COMSOL Multiphysics\textsuperscript{\textregistered}. 

\begin{figure}[t]
\centering
\includegraphics[width=\linewidth]{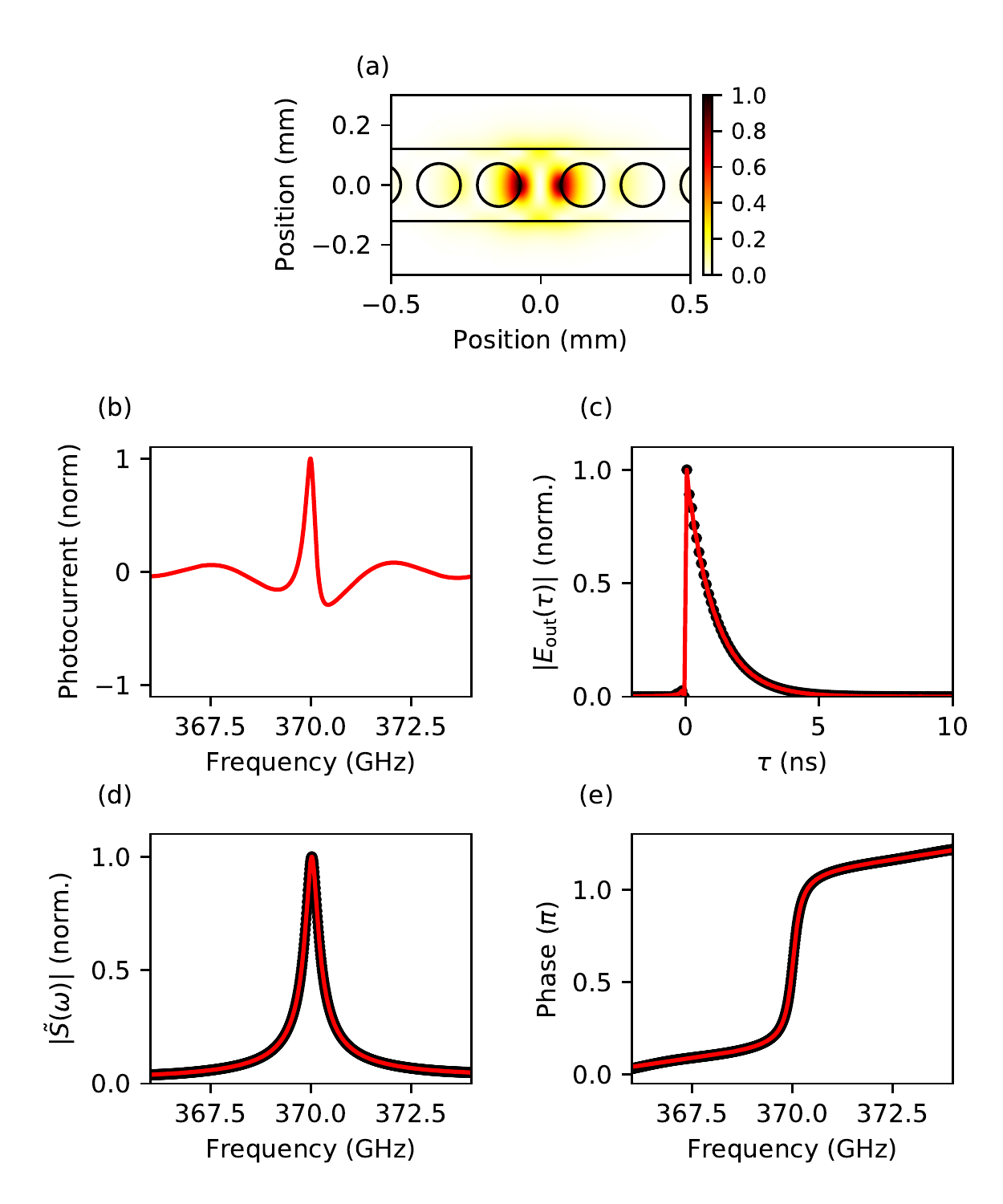}
\caption{(a) Resonant cavity mode in the PCCW (${|\textrm{E}|}^{2}$) at 370\,GHz. The silicon waveguide and the air holes are indicated with the black solid lines and circles, respectively. The width of the silicon waveguide is 240\,$\upmu$m, and the radius of the air holes is 72\,$\upmu$m. (b) Generated mock photocurrent, and (c) the corresponding envelope of the time response extracted with Hilbert transformation (black circles) and the absolute value of the complex mock data (red solid lines). (d) and (e) show the envelope and phase of the frequency response function, respectively.}
\label{fig:4}
\end{figure}

%As described above, if the mock photocurrent has a SSB spectrum, the envelopes retrieved from the complex mock data and the Hilbert transformation are indistinguishable. We base our analysis on the mock-photocurrent rather than the complex frequency response of the system, as this approach is closest to the experimental procedure.

\begin{figure}[t]
\centering
\includegraphics[width=\linewidth]{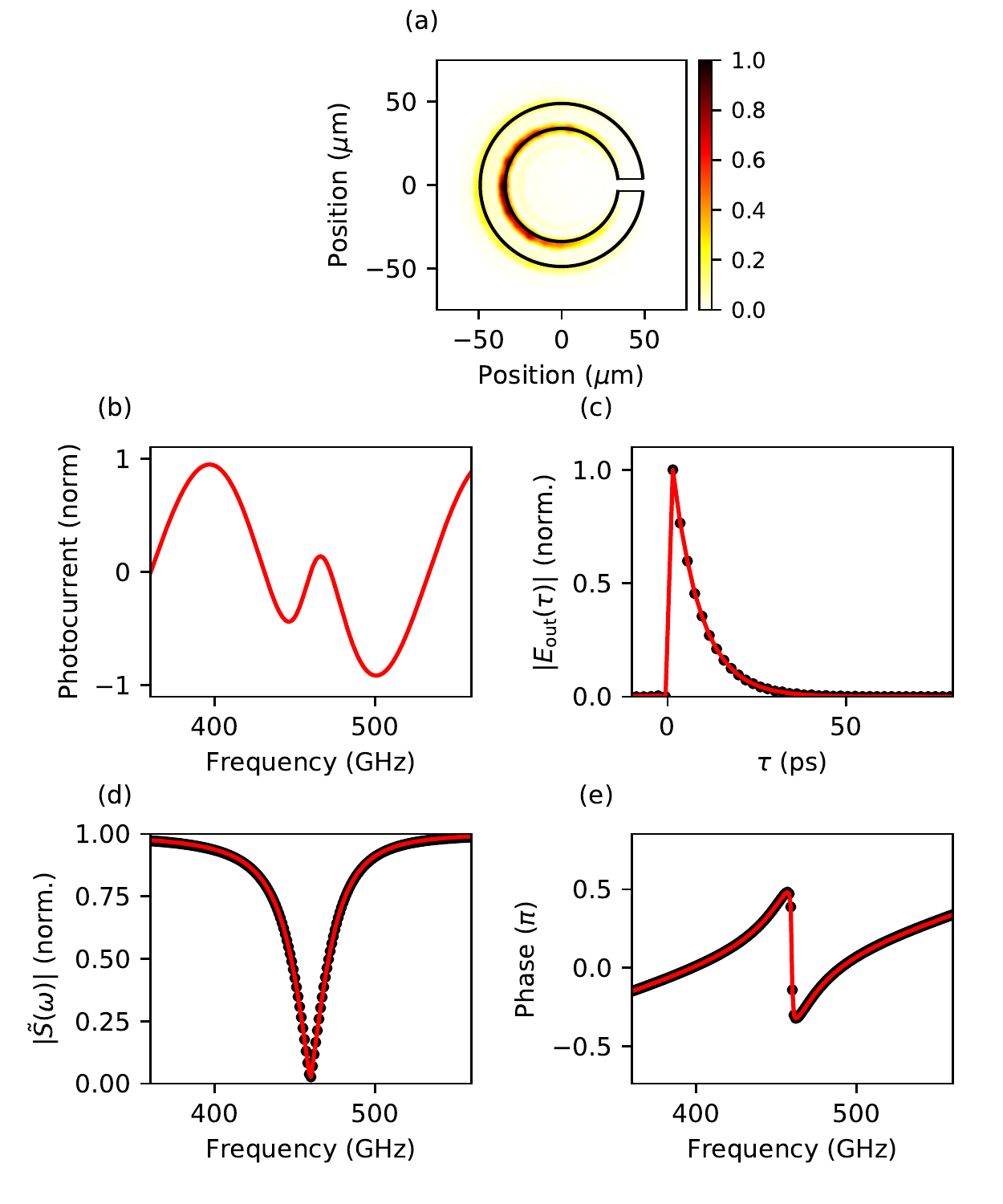}
\caption{(a) Resonant mode (${|\textrm{E}|}^{2}$) of the SRR at 460\,GHz. The SRR is indicated with the black solid lines. The outer radius of the ring is 50\,$\upmu$m, the width is 15\,$\upmu$m and the gap is 10\,$\upmu$m wide. (b) Generated mock photocurrent, and (c) the corresponding envelope of the time response extracted with Hilbert transformation (black circles) and the absolute value of the complex mock data (red solid lines). (d) and (e) show the envelope and phase of the frequency response function, respectively.}
\label{fig:5}
\end{figure}

We first consider a PCCW \cite{joannopoulos2011photonic} implemented as a single-mode silicon waveguide with a periodic sequence of eight air holes along the direction of propagation. For simplicity, silicon with a refractive index of 3.416 is assumed to be lossless \cite{vogt2018thermal}. Figure \ref{fig:4}(b) shows the generated mock photocurrent of the PCCW, using the simulated frequency response function. We can clearly observe a resonant cavity mode at a frequency of about 370\,GHz. The oscillation period of the photocurrent was here chosen to be 3\,GHz, which is about six times larger than the amplitude FWHM of the resonant cavity mode. As shown in Fig. 4(c), the time response ${E}_{\textrm{out}}(t)$ is a causal signal. Accordingly, applying the Hilbert transformation on the mock photocurrent allows for the full reconstruction of the time [Fig. \ref{fig:4}(c)] and frequency [Fig. \ref{fig:4}(d) and (e)] response functions of the system.

%The air holes form a photonic crystal (PC), leading to a broad bandgap in the transmission of the waveguide. Introducing a defect in the center of the PC leads the formation of a cavity. The high-Q resonant cavity mode trapped in the defect spectrally resides within the PC bandgap.

%Intriguingly, the corresponding envelope of the time response, obtained from the simulated frequency response is clearly a causal signal, as shown in Fig. \ref{fig:4}(c) with the red solid line. Accordingly, the envelope of the time response, as well as the envelope and instantaneous phase of the frequency response extracted with Hilbert transformation (black circles) and directly extracted from the simulated frequency response (red solid lines) are identical, as can be seen in Figs. \ref{fig:4}(c), (d) and (e), respectively. 

%Hilbert transformation applied to the mock photocurrent, shown in Fig. \ref{fig:4}(c)

Finally, we apply the same numerical procedure to a planar SRR in a periodic array. The metamaterial is designed to operate as a narrow bandpass filter at 460\,GHz. A unit cell of the metamaterial is modeled using periodic boundary conditions. The SRR is patterned in a sheet of perfect electric conductor (PEC) on top of a lossless 20\,$\upmu$m thick PTFE sheet. The spatial profile of a resonant mode at 460\,GHz is shown in Fig. \ref{fig:5}(a).

The mock photocurrent with an oscillation period of 215\,GHz, which corresponds to about nine times the amplitude FWHM of the resonant mode is shown in Fig. \ref{fig:5}(b). Please note that the mock photocurrent is generated assuming a reflection measurement, and not transmission as for the PCCW above. The envelopes of the time responses obtained from the simulated frequency response and from the Hilbert transformation applied to the mock photocurrent are shown in Fig. \ref{fig:5}(c), confirming the causality of the system. Again, a perfect agreement between the envelope [Fig. \ref{fig:5}(d)] and instantaneous phase [Fig. \ref{fig:5}(e)] extracted with Hilbert transformation (black circles) and the absolute value of the simulated frequency response function (red solid line) is observed. Interestingly, a very similar envelope and phase profiles have been previously experimentally observed for more sophisticated split ring resonators \cite{chen2006active,chen2009metamaterial,5612801}.

We have shown that, in the context of coherent CW THz spectroscopy, Hilbert transformation enables full frequency response analysis of physical linear systems regardless of the widths of the resonance features under study. This observation appears to be in contrast to the general conception that the Hilbert transformation is only applicable to signals with a slowly varying envelope compared to the underlying oscillation period. However, because the photocurrent measured in typical CW THz spectroscopy experiments is the real part of an analytical signal, Hilbert transformation is applicable regardless of the oscillation period and the specific modulation of the photocurrent. Ultimately, our results show that the unrestricted applicability of the Hilbert transformation ensues from the causality of the investigated linear systems. The presented results provide extremely powerful tools for CW THz spectroscopy, in particular for systems with narrow resonance features. More generally, our 
work highlights how concepts from linear systems' theory and complex functions can enable breakthrough advances in contemporary technologies.

\begin{acknowledgments}
M. Erkintalo acknowledges support from the Rutherford Discovery Fellowships of the Royal Society of New Zealand.
\end{acknowledgments}

\end{document}